\documentclass[%
 reprint,
superscriptaddress,
 amsmath,amssymb,
 aps,
floatfix,
]{revtex4-2}

\usepackage{xcolor}

\usepackage{graphicx}
\usepackage{dcolumn}
\usepackage{bm}

\begin{document}

\preprint{APS/123-QED}

\title{Thermal Schwinger Effect: Defect Production in Compressed Filament Bundles}

\author{Valentin M. Slepukhin}
\affiliation{Department of Physics and Astronomy, UCLA, Los Angeles California, 90095-1596, USA}
 
\author{Alex J. Levine}%
\affiliation{Department of Physics and Astronomy, UCLA, Los Angeles California, 90095-1596, USA}
\affiliation{Department of Chemistry and Biochemistry, UCLA, Los Angeles California, 90095-1596, USA}

\date{\today}

\begin{abstract}
We discuss the response of biopolymer filament bundles bound by transient cross linkers to 
compressive loading.  These systems admit a mechanical instability at stresses typically below that of 
traditional Euler buckling.  In this instability, there is thermally-activated pair production of 
topological defects that generate localized regions of bending -- kinks.  These kinks shorten the 
bundle's effective length thereby reducing the elastic energy of the mechanically loaded structure.  
This effect is the thermal analog of the Schwinger effect, in which a sufficiently large electric 
field causes electron-positron pair production. We discuss this analogy and describe the implications of 
this analysis for the mechanics of biopolymer filament bundles of various types under compression.
\end{abstract}

\maketitle


Long, stiff filaments held together by strong bonds are ubiquitous in biology.  These
filaments appear in both the cytoskeleton and the extracellular matrix in the form of bundles 
bound by a variety of cross-linking molecules, which, due to their weaker interactions with the  
filaments, attach and detach from the bundle reaching a chemical equilibrium with their 
concentration in the surrounding fluid.  The mechanical response of filaments and their networks 
is well understood.  The filaments are nearly inextensible; they respond
to tensile or compressive loading by reducing or increasing (respectively) the amount of 
filament arc length stored in their transverse thermal
undulations~\cite{mackintosh1995elasticity,Morse:1998,Maggs:1999,Broedersz:2014}.

\begin{figure*}
    \centering
    \includegraphics[width=\linewidth]{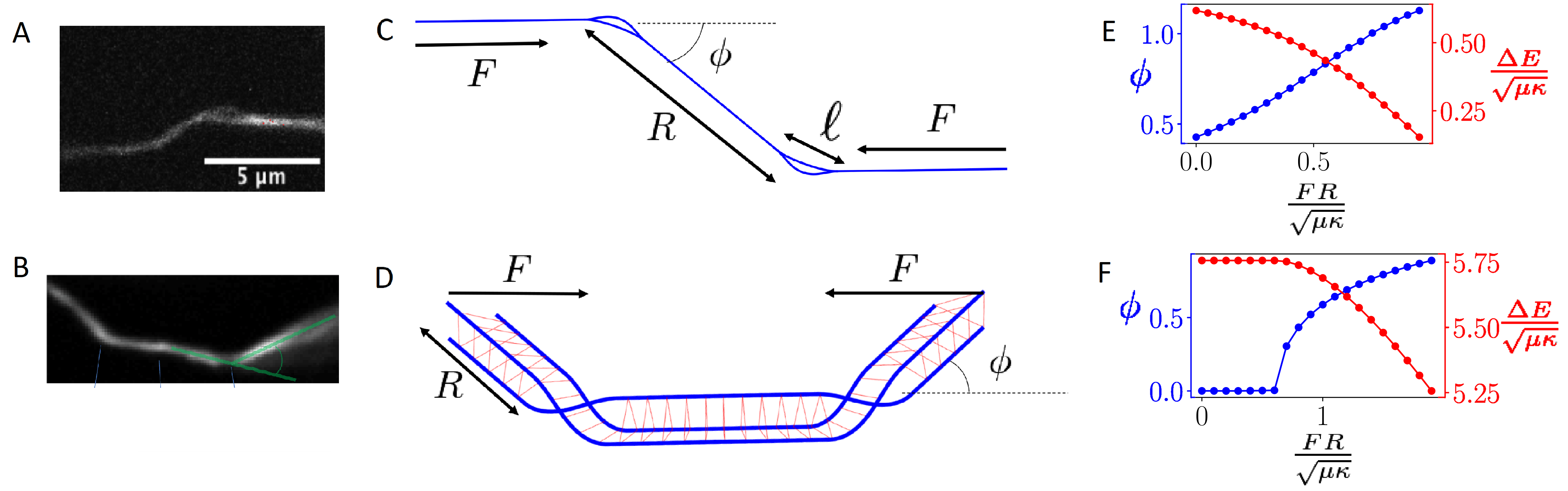}
    \caption{Fluorescence microscopy images of a z-bend (A) and a u-bend (B) in a collagen bundle. (C) Two loops  
    under compression form a z-bend. (D) Two braids under compression form a u-bend. (E) Angle $\phi$ produced by 
    a loop pair (blue, left axis) and energy difference between the looped and straight bundle as a function of dimensionless torque 
    (red, right axis). (F) Angle $\phi$ produced by a braid pair (blue, left axis) and the energy difference 
    between braided and straight bundles as the function of dimensionless torque (red,right axis). 
    [Images courtesy of E.~Botvinick and Q.~Hu] }
    \label{fig:schematic}
\end{figure*}
The collective mechanical response of filament bundles is more complex than that of simple 
elastic rods.  For instance, bundles have a length-scale dependent bending modulus~\cite{Frey:07} 
whereas the underlying filaments typically do not.  The increased thickness of the bundle 
suppresses collective bending deformations, so bundles are significantly less compliant 
than their constituent filaments. But bundles admit new internal degrees of freedom associated with 
the filament reorganization. In particular, there are two relevant defects 
associated with these rearrangements: braids, corresponding to the transposition of filaments within the 
bundle, and loops, which trap extra length in a subset of the bundle's 
filaments between consecutive cross links~\cite{Slepukhin:2021}. 
See Fig.~\ref{fig:schematic}C and~D for schematic diagrams of pairs 
of loops and braids respectively.  Forming these defects from a quench by adding cross linkers is 
commonplace, but since the addition or removal of these defects requires a system-sized rearrangement of 
cross linkers, one cannot expect them to form spontaneously.  Rather, they form in defect -- anti-defect 
pairs, which require only local cross-linker rearrangements. Since these defect 
pairs are associated with kinks, compressive loading suppresses the energy barrier associated 
with defect pair production.  As a result, at a critical compressive stress, we 
expect the proliferation of these defect pairs once the energy cost of pair production is reduced to the 
thermal energy. 

Stiff rods under compression are known to undergo a mechanical instability called  Euler buckling. 
Upon compression directed along its long axis, an elastic rod first 
shortens its length, bearing the external loading via compression. At a critical strain, the rod 
buckles, supporting the compressive stress via bending on the length scale of the 
entire rod~\cite{Landau1986,Golubovic:1998}.  When we consider the response to compressive 
loading of biopolymer filament bundles bound by transient cross linkers, we find that 
Euler buckling is precluded in such composite objects at finite temperature by another type 
of instability: at compressive stresses lower than the
Euler buckling threshold, the bundles shorten by the thermally-activated production of pairs of
topological defects, leading to localized regions of bending deformation -- {\em kinks} -- 
unlike the system-sized bends encountered in Euler buckling.

Defect pair production is analogous to the Schwinger effect, in which electron-positron pair 
production was predicted in a sufficiently strong static 
electric field~\cite{schwinger1951gauge,schwinger1954theory}.  
The forces due to the large electric field on the charged particle pairs 
pulls them apart, stabilizing these quantum fluctuations of the 
vacuum. In the same way, it is energetically favorable for thermally-generated defects to be produced
under bundle compression.  We term this mechanism the {\em thermal Schwinger effect}. 

To make this analogy more precise we elaborate briefly on the correspondence.  The 
stored length in a loop (see Fig.~\ref{fig:schematic}C) $ \Delta l$  corresponds to 
a continuously variable electric charge.  In analogy to charge conservation, 
loop defects must be formed to conserve the total stored length so that all the
filaments remain in registry to the left and right of the defect pair.  Otherwise defect pair production would
require a system-sized reorganization of all cross linkers. 
For the case of braid defects, each defect (for a three filament bundle) is associated with an elementary
operation of the modified braid group~\cite{Artin:1947} $Br_3 \times Z_2$ modulo rotations of the entire 
bundle -- see SI. These defects also store length and thus carry the scalar 
charge associated with loops as well as a charge associated with $Br_3 \times Z_2$.  The 
braid group charge is quantized, but there will be a continuous spectrum of length above a gap set by the minimum stored
length necessary to produce the braid. In fact, the most general form of a defect is one that carries both scalar and 
braid-group charges.  In order to simplify our analysis, loops refer to defects with zero braid-group charge, and
braids carry the scalar length charge necessary to minimize their defect core energy. We leave more 
complex defect structures to future work.  To preserve the cross-linking structure far away from the defect 
pair, the net braid charge of the defects must vanish, i.e., the braid/anti-braid pair must be associated with inverse 
operations of the braid group. The end-to-end length of the bundle is conjugate to the applied compression.  
Defect/anti-defect pair production shortens the end-to-end distance of the bundle both through length stored in the
defects cores and the kinking of the bundle at the defects and thus reduces the energy cost of the formation of 
the defect pair. This is analogous to how an applied electric field decreases the 
energy cost of electron/positron pair production in the Schwinger effect.  For the case of current interest, however, the
defect pairs are created by thermal, not quantum, fluctuations. Loop defects separate
under compression, like the electron/positron pairs in the applied electric field.  Braids, however, attract each other 
forming bound states~\footnote{Braids also repel under tension, see~\cite{slepukhin2021oscillations}}.  
This attraction has no analog in the standard Schwinger effect. 

We first consider the energetic cost of the production of loop and braid 
defects in a compressed bundle. We then use these results to compute the loop pair production rate at temperature 
$T$ in a calculation reminiscent of the Kramers' escape problem~\cite{kramers1940brownian, vankampen2007spp}.
We also analyze the critical stress for defect pair production in a few biopolymer systems including 
F-actin, DNA, and collagen bundles. 

To compute the minimal energy configuration for stable and metastable states of the $N$-filament bundle under a 
compressive force $F$, we introduce the energy 
\begin{equation}
E = - F \Delta L + \mu \ell + \sum_{i=1}^N  \int ds  \frac{\kappa_i}{2} \left(\partial_s \hat{t}_i \right)^2.
\label{eq:general-energy}
\end{equation}
The first term gives the energy reduction due to the shortening of the bundle's end-to-end distance $\Delta L$. 
The cross linkers have binding energy $\mu$ per unit length. Since defects 
disrupt cross linking over a distance $\ell$,  their presence increases the system's energy as reflected by 
the second term on the right hand side of Eq.~\ref{eq:general-energy}.  The third term gives 
the bending energy stored in the bundle, where $\kappa_i$ and $\hat{t}_i(s)$ are the bending modulus and tangent 
vector of the $i^{\rm th}$ filament. $s$ is the arc length along the bundle.  We neglect torsion, 
so all defect energies are actually lower bounds.  There will be a continuous
spectrum of excited states due to trapped torsion.

We examine first a pair of loop defects while assuming 
the compressive load to be sufficiently weak so the characteristics of the loop, i.e., the dependence 
of its kink angle and energy on its size can be taken from our previous 
calculations in the zero-compression limit~\cite{Slepukhin:2021}. We discuss the effect of the angle change later 
and in the SI.  The kink angles generated 
by neighboring loops are equal and opposite, since the amounts of their trapped length have to be equal and 
opposite (which also makes the loop sizes equal, see SI).  
A pair of loops produce a {\em z-bend} where  parts of the bundle not lying between the loop pair are 
parallel and offset in the normal direction to the undeformed bundle -- Fig.~\ref{fig:schematic}A,C. 
This result holds even for bundles having filaments of differing bending moduli, as long as 
the excess trapped length in the loop is much smaller than the total length of the defected region.
For simplicity, we focus on the case of equal bending moduli. Then the total energy of 
configuration with two loops of size $\ell / 2$ each, generating kink angles $\phi$, and 
separated by a distance $R$ is
\begin{equation}
    E_{\rm tot} =  g_1 \mu \ell   - F R ( 1 - \cos \phi). 
    \label{eq:tot-en}
\end{equation}
The first term in the Eq.~\ref{eq:tot-en} is the energy of the pair of loops of length $\ell / 2$, 
with  coefficient $g_1 \approx 1.48$ (see SI). The second term is the decrease of energy due to the 
compression (see Fig.~\ref{fig:schematic}C). As long as $F \ll \mu$, it is not important whether we 
define $R$ to be the distance between centers of loops or their edges, since the difference will be small in 
comparison with the first term. However, we pick $R$ to be the distance between closest edges, 
so it is equal to zero when loops are not yet separated.

There is a continuous distribution of loop sizes, leading to a continuous distribution of angles of the
z-bends produced by loop pairs and a similar distribution of energy reductions associated with them.  
Observed loop pairs are the result of a stochastic process
of pair production, which is related to the classic problem of the thermally-activated escape from 
a potential well.

We investigate the energetics of pair production and escape.  Loop formation 
involves cross linker removal and filament bending, leading to an energy increase of $g_1 \mu \ell$ as the loop 
size $\ell$ increases. At some loop size $\ell_0$, the two
growing loops separate due to random fluctuations.  Once separated, the loops can no longer exchange 
trapped length so their lengths are now fixed at $\ell_0/2$ each (see SI). 
As the distance $R$ between the loops of the resulting z-bend grows, the energy of the compressed bundle decreases
due to shortening along the direction of the compressive load. We can consider this process as an escape 
from the potential well using $x$ as a single reaction coordinate that describes the growth of the loop 
size while they overlap and then their separation afterwards:
\begin{equation}
    U(x) = 
    \begin{cases}
    g_1 \mu x , x < \ell_0 \\
    g_1 \mu \ell_0 -  F  ( 1 - \cos \phi) (x - \ell_0), x > \ell_0,
    \end{cases}
    \label{eq:effective-potential}
\end{equation}
$x$ grows with the sizes of the loops $x = \ell$ before separation (upper equality), and then
describes the distance between the separated loops $x = R$ (lower equality). The effective potential for the 
growing loops increases linearly with loop size up to the final loop size
$\ell_0$  and then decreases linearly 
due to the shortening of bundle along the direction of the applied force. 
Taking into account the change of the angle due to increasing torque leads to faster decrease of the potential, which 
accelerates pair production. Here we present the calculation for the lower limit of the production rate, 
when the angle is assumed to be 
constant, and we analyze the effect of the angle change in the SI.
\begin{figure}
    \centering
    \includegraphics[width=\linewidth]{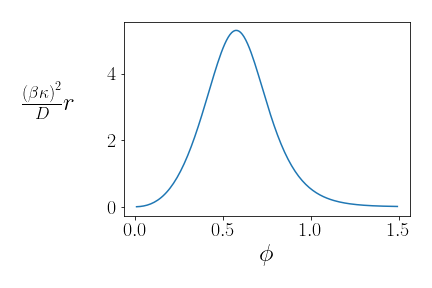}
    \caption{ Dimensionless loop pair production rate with 
    $\eta = 10$, $\tau = 0.1$ (see Eq.~\ref{eq:rate})}
    \label{fig:kramers}
\end{figure}

Treating pair production as a Kramers' escape problem~\cite{kramers1940brownian} in the 
potential Eq.~\ref{eq:effective-potential}, we compute the escape rate $r$, 
the rate of loop pair production in thermal equilibrium at a fixed compressive stress. We 
compute this rate as the inverse of the mean time to escape using the standard Kramers' 
approach for an overdamped system: 
\begin{equation}
    r^{-1} = \frac{1}{D}  \int_0^{x_0} dy e^{\beta U(y)} \int_0^y dz e^{-\beta U(z)},
    \label{eq:kramers}
\end{equation}
where $x_0$ is defined such that $U(x_0) = 0$ and $\beta = 1/k_{\textrm{B}}T$.
The reptative motion of the defects is heavily overdamped; we introduce a loop diffusion constant 
$D \propto k_{\textrm{off}} \Delta x^2$ in terms of $k_{\textrm{off}}$ the rate of 
cross-linker unbinding and the distance between consecutive binding sites of those cross linkers along the 
filament $\Delta x$.

Performing the integral in Eq.~\ref{eq:kramers} in the limit of small $F / \mu $ and $\phi$ (see SI), 
we obtain
\begin{equation}
r^{-1} = \frac{4}{D \beta^2 F^2 \phi^4 } \left(  \frac{ \tau \phi^2 }{2 g_1  }     e^{  \eta \phi }       +
      \eta \phi    -  1   + e^{- \eta \phi } \right)  ,
\label{eq:rate}
\end{equation}
where we introduce the dimensionless parameters $\tau = F / \mu$, $\eta  = \ g_1 g_2 \beta  \sqrt{\kappa  \mu}$, with
$g_2 \approx 4.8$ relating defect size to the kink angle it produces: 
$\ell = g_2 \sqrt{\frac{\kappa}{\mu}} \phi$ (see  SI).
The pair production rate $r$ vanishes as $\phi$ goes to zero since the potential barrier width 
diverges as $1/\phi^2$.   Conversely, very large angle kink production is also suppressed 
($r  \rightarrow 0$ as $\phi \rightarrow \infty$) due to the increasing 
energy of the loop. The rate of pair production has a maximum at a finite angle -- see 
Fig.~\ref{fig:kramers}.  We obtain a prediction for the most commonly produced 
kink angles in z-bends as a function of material parameters of the bundle and the applied compressive load. 
In the limit of weak compression, the maximum loop pair production rate 
$r_{\textrm{max}} $  (z-bend formation rate) 
occurs at angle $\phi^\star$  (see SI for details):
\begin{eqnarray}
\phi^\star &=& \eta^{-1} \log \left(\frac{6 g_1 \eta^2}{\tau} \right)\\ 
r_{\textrm{max}}  &=&  
\frac{D}{3} \left[ \frac{ \tau   \log \left(\frac{6 g_1 \eta^2}{\tau} \right)  }{2  \beta \kappa  } \right]^2 
\end{eqnarray}
The production rate of the 
z-bends increases as the compressive force squared and is rather sharply peaked -- 
Fig.~\ref{fig:kramers} -- as a function of angle, suggesting that, for 
fixed material parameters, including bundle sizes, one expects to observe a narrow range of z-bending angles. 
The most probable z-bend angle scales roughly as $k_{\textrm{B}}T/\sqrt{\kappa \mu}$; the binding energy 
of the linkers determines the typical observed angles for bundles of a fixed number of filaments.  Finally, 
as the bundle size grows, the effective $\kappa$ increases, driving the z-bend angles to zero.

We now examine the production of braid/anti-braid pairs in a three-filament system.  Within the lowest 
energy configuration of the braid, two of the filaments follow the same trajectory, allowing us to reduce the
problem to that of studying two filaments in 2D.  
We call the case of two filaments with equal bending moduli a pseudobraid, reserving the name 
braid for the more physical but analytically less tractable case unequal bending moduli $\kappa_1 = 2 \kappa_2$.
See Ref.~\cite{Slepukhin:2021} for further details. 

Unlike in the case of loops, 
only the magnitude of the kink angles produced by the braid pair must be equal.  
The kink angles generated by braids thus do not have to form z-bends; in fact, the 
lowest energy state will be 
a {\em u-bend} as shown in Fig.~\ref{fig:schematic}B. This energy is minimized when the 
two defects are close to each other and localized in the middle of the bundle, 
since this provides the greatest 
shortening in response to the force. We speculate that braid defect co-localization is the 
primary reason for the rarity of u-bend observations as compared to z-bends 
(see Ref.~\cite{Slepukhin:2021}).  U-bends could be easily misinterpreted as a 
single defect with a larger kink angle. 

In minimizing the total energy of the bundle (see SI), it is convenient to introduce a dimensionless 
parameter $\zeta = \frac{\mu a^2}{\kappa}$, where $a$ is the spacing between the centerlines of the 
filaments enforced by the cross linkers.  
We expect this distance
to be the sum of the linker size and twice the radius of the filament's cross section.  Using the same 
parameters, we also introduce a dimensionless applied force $f = \frac{F R a}{\kappa}$. 
We find that, up to a critical compression $f^\star(\zeta)$, implicitly determined by
\begin{equation}
\int_0^1  \frac{\frac{  ( \zeta  -  \sqrt{2 \zeta })^2 }{   f^2 } t dt}
{\sqrt{1 - \frac{  ( \zeta  -   \sqrt{2 \zeta })^2 }{   f^2 } t^2} \sqrt{1  - t}  }  = \sqrt{\zeta/2},
\label{threshold-equation}
\end{equation}
the minimum energy configuration of the braid/antibraid
pair remains that of an unkinked bundle as shown in Fig.~\ref{fig:schematic}F.  
This is distinct from the case of loop pairs where low-angle loops can form 
at any compressive load. 
For $f > f^\star (\zeta)$, the defect pairs produce finite-angle kinks 
-- Fig.~\ref{fig:schematic}F -- resulting in a u-bend whose angles grow with $f$.

\begin{figure}
    \centering
    \includegraphics[width=\linewidth]{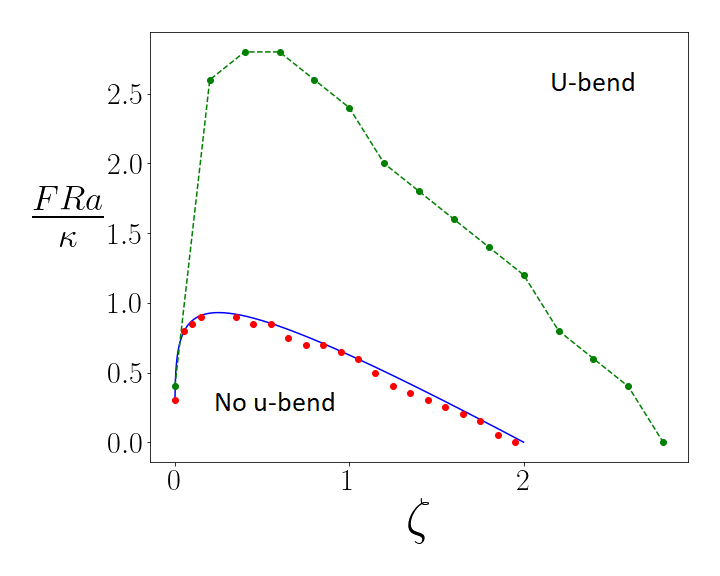}
    \caption{Numerical minimization of the energy (red dots) and analytical prediction implicitly given by Eq.~\ref{threshold-equation} 
    for a symmetric pseudobraid $\kappa_1 =\kappa_2$. The 
    numerical solution of the more complex, three-filament braid with $\kappa_1 = 2 \kappa_2$ (green circles, 
    green dashed line is a guide to the eye) shows
    that the transition is shifted to higher compression.}
    \label{fig:threshold}
\end{figure}

Solving Eq.~\ref{threshold-equation} numerically (which agrees with the 
numerical minimization of the energy Eq.~\ref{eq:general-energy}), we obtain a phase diagram 
spanned by compressive loading $f$ and $\zeta$ shown in 
Fig.~\ref{fig:threshold}. Above and to the right of the boundary, u-bends are present.  One can 
interpret the diagram as a graph of the critical loading versus linker binding energy $\mu$ 
at fixed $\kappa$ and $a$. The non-monotonic behavior of the curve can be understood as follows.  
For sufficiently large $\mu$, kinks appear at braids even at zero compressive stress, but as the linker 
binding energy decreases, kink formation is energetically unfavorable unless the shortening of the bundle under 
load produces a sufficient energy reduction. For small enough
linker binding energy, the defected regions extend in arc length, thereby becoming more bending compliant so that 
there is a re-entrant kinking regime at small $\mu$.   The behavior of the more physical, asymmetric case (green 
circles) is similar to that of the pseudobraid (red circles and blue line), but the transition is shifted to 
higher compressive loads due to the increased bending rigidity of the system. 

Upon increasing the compressive load, we predict that bundles should first shorten by producing loop pair 
defects creating z-bends, as found in the collagen bundles seen in Fig.~\ref{fig:schematic}A. Assuming the 
size of the bundles is known and controlled, the resulting z-bends will be generated with reproducible angles,
due to the peak in stochastic defect production rate with angle as shown in Fig.~\ref{fig:kramers}. 
We expect the high polydispersity of typical biopolymer filament bundles to spread out the distribution of 
z-bend angles. But since the angle of maximum production 
$\phi^\star \sim k_{\rm B} T/\sqrt{\kappa \mu}$ and for 
a bundle of $N$ filaments $\kappa \sim N^2$, we expect $\delta \phi^\star \sim \delta N N^{-2}$. The
peak in the z-bend angle distribution may be hard to observe without some bundle control 
unless $N$ is large.  If the cross-linking energy is sufficiently
large, the z-bend angles will vanish as $\phi \sim 1/\mu^{1/2}$. However, as the distance $R$ 
between the two loops increases, we cannot continue to neglect the increase of the equilibrium loop angle 
shown in Fig.~\ref{fig:schematic}E, which may lead to observable angles at large $R$, even if they were
unobservably small angles at formation.  
As shown in the SI, loop pairs will generically deform to sharp angles -- crumple -- as they separate. 
Such large angles have not been observed in collagen bundles~\cite{Slepukhin:2021}. This may be due to 
one of two possibilities. The
bundles may be short enough that z-bends would have to diffuse off the ends to reach 
sufficient torque for crumpling,
or defect motion may be so slow that their equilibrium state is not typically observed.

At higher compression, the u-bends seen in Fig.~\ref{fig:schematic}B 
will also be created when braid pair production is reduced to thermal energy. 
One may ask whether braids or loops are preferentially generated under particular 
conditions of fixed torque.  
Defect formation is an inherently stochastic process, but we expect that, since there is a continuous 
spectrum of low-energy, small stored length loops, these should form preferentially at lower temperatures.  To 
further examine this point, we provide in the SI a phase diagram showing that loops storing small amounts
of excess length leading to smaller kink angles $\phi$ are energetically favored over braid pair production 
when the loop kink angles remain below a threshold $\phi(\zeta)$. 

Using estimates of $\zeta$ for various filament
bundle systems, we find F-actin and collagen to have $\zeta \sim 0.1$~\cite{Slepukhin:2021}, 
so uncompressed bundles support
unkinked braids. Braid pair production leading to u-bend formation should occur for compressive 
forces on the order of 10 pN based on the phase diagram shown in Fig.~\ref{fig:threshold}.  
DNA condensed by polyvalent ions and cross-linked intermediate 
filaments have $\zeta \sim 100$~\cite{Slepukhin:2021} 
suggesting that there will be a number of 
kinked braids quenched into the bundle.  As a result, we expect these bundles to collapse by bending at 
the preexisting braids, which introduce more bending compliant regions via cross linker reduction. Finally, we 
note that under sufficiently large forces, Euler buckling can take over from braid-generated u-bend formation.  
We estimate that Euler buckling should be found for $F R a/\kappa \approx 5$ (see SI) for 
$\zeta \approx 0.1$ (and this value only grows for larger $\zeta$) which is well above the region shown in 
the u-bend phase diagram, Fig.~\ref{fig:threshold}.

The most direct test of the theory should be found in compression experiments on 
individual bundles.  For collagen and F-actin, the necessary compressive forces are on the order of 
10pN, suggesting laser trapping experiments should probe the relevant force scales. One of the implications 
of defect pair production is that they introduce a way to maintain bundle integrity under 
large compressive loads by providing a means for local failure of cross linking without the global breakdown of the bundle itself.
There remain a number of open questions about more complex defects and their 
interactions on the bundle.  Firstly, more complex defects containing excess length and braids may form and have interesting dynamics
in that they may exchange length via the transport of pure loop defects between them. Because the spectrum of 
energy and excess length in loop defects is continuous, complex defects may exchange low-energy loops or arbitrarily small
excess length in their relaxational dynamics. Secondly, given that bundles in a biological context are often found in conjunction 
with molecular motors, one may ask how motor-induced forces drive defect dynamics.

AJL and VMS acknowledge partial support from NSF-DMR-1709785 and thank the Botvinick group (UCI) for 
sharing unpublished collagen data and for illuminating discussions.  
VMS acknowledges support from the Bhaumik Institute Graduate Fellowship and Dissertation Year Fellowship, UCLA.

\appendix

\section{Loop pair production as a Kramers' escape problem}

In the following we provide detailed calculations of the loop generation problem using the formalism of the
Kramers' escape problem in one dimension.

\subsection{Forming a z-bend: The energetics of a loop pair production}
We considered the energetics of a loop in the absence of the external force in Ref.~\cite{Slepukhin:2021}. There 
we found that, for a particular amount of excess trapped length $\Delta l$, the angle of the loop 
that minimizes its energy is 
\begin{equation}
\label{eq:angle-trapped-length}
\phi  = 2  \frac{ \Delta l ^{1/3}  \mu^{1/6} }{\kappa^{1/6} x^{1/3}  }    \tanh (x),
\end{equation}
where $x \approx 2.365$ is a first non-zero solution of the equation $\tan x =  - \tanh x$. We 
can also express this minimum loop energy $E$ and its size $L$ 
(the length of the shorter filament(s) that is 
not cross linked to the looping filament) as the function of the trapped length:
\begin{equation}
\label{eq:energy-trapped-length}
E =  3 \mu^{2/3}   \Delta  l^{1/3}  \kappa^{1/3}  x^{2/3}     
\end{equation}
and
\begin{equation}
\label{eq:length-trapped-length}
L    =    \Delta  l^{1/3}  \frac{ 2 \kappa^{1/3}   x^{2/3} }{ \mu^{1/3} }.
\end{equation}
From Eqs.~\ref{eq:angle-trapped-length},\ref{eq:energy-trapped-length}, 
and \ref{eq:length-trapped-length} we observe that
amount of the trapped excess length $\Delta l$ fully controls the kinking angle $\phi$, the energy of the defect
$E$ and the length of the defected region $L$. As a result, if the trapped length is equal and opposite for 
two loops (the excess length on one set of filaments in the first defect compensates for the length deficit
on the same filaments in the second defect), the angles $\phi$ of the two defects will also be 
equal and opposite. Similarly, the loop sizes $L$ and energies $E$ of the two defects will be equal.

If the total size of the uncross-linked region is $\ell$ (including both loop defects), then $L = \ell /2$, 
and we can express $\phi$ and $E$ in terms of $\ell$:
\begin{equation}
\phi  =  \frac{\ell \mu^{1/2} }{2\kappa^{1/2}x }\tanh (x) = \frac{1}{g_2} \sqrt{\frac{\mu}{\kappa}} \ell
\label{eq:angle-size}
\end{equation}
and 
\begin{equation}
E =  3 \mu \frac{\ell}{ 4} \tanh (x)^{2/3}  = g_1 \mu \ell / 2.     
\end{equation}
These are easily obtained from Eqs.~\ref{eq:angle-trapped-length} and \ref{eq:energy-trapped-length} above using 
$\ell$  appropriately. This pair of length-compensating loops together constitute a single z-bend, as discussed
in the main text. 

If we now assume that the compression force $F$ is small enough, we may ignore the finite-force 
corrections to the angle. Then, for a given compressive load $F$ acting on a bundle containing a pair of 
length-compensating loop defects that produce a z-bend where a length $R$ separates the two loops, the 
decrease of the bundle's energy 
is the work done by the compressive load in shortening it: $-F R(1 - \cos \phi)$.
Taking this into account, we obtain the total energy of the z-bend as the sum of this work and energy 
cost of the formation of the two defects
\begin{equation}
\label{eq:z-bend-energy}
E_{tot} =  g_1 \mu   \ell   -     F R (1 - \cos \phi).  
\end{equation}
As long as $F \ll \mu$, changes in $R$ of the order of $\ell$ produce small changes to the energy when compared
to the first term in Eq.~\ref{eq:z-bend-energy}. Thus, in this limit it does not matter whether we define 
$R$ to be the distance between the defect centers or the defect edges. For consistency, we consider $R$ to be the 
distance between the closest edges of the defects, i.e., when the defects are starting to separate, 
$R$ grows from 0.

\subsection{The effect of variations in the kink angle and bundle collapse}

\begin{figure*}
    \centering
    \includegraphics[width=\linewidth]{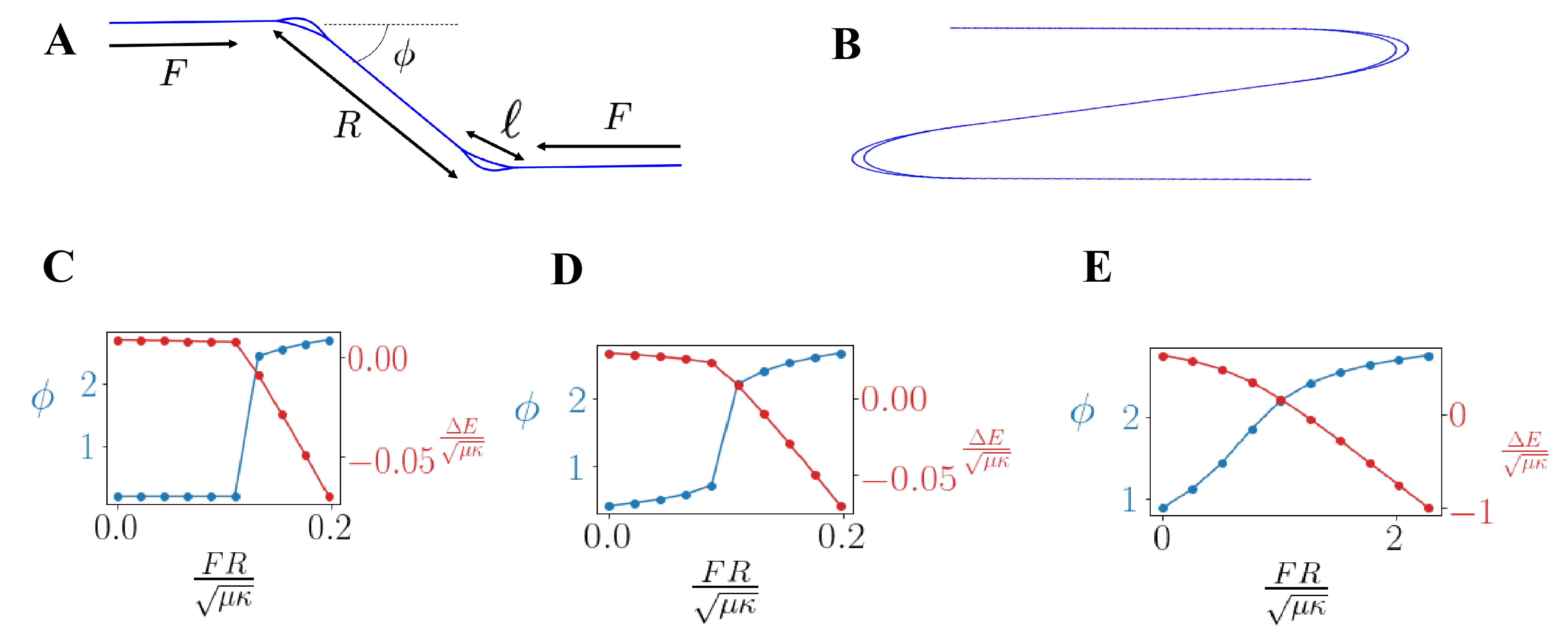}
    \caption{(A) Smooth and sharp (B) z-bend respectively.  (C) The kink angle is small and almost 
    constant ($\approx$ 0.1 rad) until the threshold value of torque where the bundle crumples. (D) The kink 
angle first increases smoothly and then jumps discontinuously at the threshold. (E) The kink angle 
increases continuously with increasing torque.}
    \label{fig:schematic-si}
\end{figure*}

As we increase the torque, either by increasing the force or by increasing the moment arm, the kink angle should
increase as well. Here we consider the validity of our former approximation (see the main text and above) that the
kink angle is essentially constant and explore what will happen when the kink varies.  
As we see from the numerical simulation, the exact sequence of angular changes depends on such parameters 
as the amount of trapped length and cross link size (unlike analytical calculations, 
the numerical simulation allows for non-zero cross link size, thus making it more realistic). In summary,
we find that there are two qualitatively distinct behaviors: (1) the kink angle remains 
approximately constant as the torque increases until a particular 
threshold value, at which point it then abruptly jumps to large values (crumpling of the bundle);  or (2) 
the angle grows continuously with increasing torque so that there is no distinct ``snap through'' or 
crumpling event. See Fig.~\ref{fig:schematic-si} for diagrams of a smooth z-bend corresponding to 
an uncrumpled bundle (A) as shown in the main 
text and a sharp z-bend (B), which occurs after a snap through or crumpling event.   

The principal effect of the changing kink angle is that, as the kink pair separates, its energy decreases faster
than one would infer from the constant angle approximation.  This is
reasonable in that one expects that, given more degrees of freedom via an adjustable kink angle, the 
total energy of the defect pair is further reduced.  This energy reduction 
depends on the inter-kink separation.  As a result, the z-bend production rate calculated in the main text 
is a lower limit of the physical rate.  Reductions in the energy cost of defect production due to changes 
in the kink angles enhance their thermal production. 

To simplify the analytical calculation, instead of taking into account a 
continuous increase of the angle with 
the torque, we consider only the case when the angle is constant until a threshold value of the torque, 
and then abruptly increases. This is a correct description for the scenario with the small amount of 
trapped length, where the angle is almost constant before the threshold. Varying the 
threshold, we can go in between two limiting cases: (1) the constant angle approximation that gives 
the lower estimate for the production rate, and (2) the case where the angle jumps to a larger value 
associated with crumpling almost immediately after the braid formation, which will give the highest 
estimate of the production rate.

\subsection{The thermal production of loop pairs as a Kramers' escape process}

We now consider the process of z-bend formation as the thermally activated production of two loops.  
This process has two essentially distinct steps.  In the first, two loops are created on the bundle. 
To conserve the length of filaments involved, the excess trapped length in one loop is 
compensated by an equal amount of trapped length in 
the other loop, but now associated with the other filament.  In short, one can look at this process as the 
exchange of a conserved amount of length trapped between cross links in two overlapping structures, forming a
loop pair.  In the second part of this process, the two loops diffusively separate to form the z-bend. During 
the first part of this process, the two loops can easily continue to exchange length so that the trapped 
length within in the loops changes.  Once the loops separate, such exchanges of length would require the 
reconfiguration of all cross links in the bundle section between the two loops.  This is energetically
prohibitive so the loop sizes are now fixed.  

These two distinct processes can be combined into a single one using a particular reaction coordinate, which 
represents the changing amount of trapped length in the first part and then the inter-loop separation in the 
second part.  The effect potential associated with this single reaction coordinate can be written as follows (Eq.~\ref{eq:effective-potential}):
\begin{equation}
    U(x) = 
    \begin{cases}
    g_1 \mu x , x < \ell_0 \\
    g_1 \mu \ell_0 -  F  ( 1 - \cos \phi) (x - \ell_0), x > \ell_0.
    \end{cases}
\end{equation}
In the following it is helpful to define $A = g_1 \mu$ and $B = F  (1 - \cos \phi)$. Then, for a loop pair 
containing excess length equal to $\ell_0$, the effective potential for the reaction coordinate may be written 
simply as an initially linearly increasing function with slope $A$ connected to a linearly decreasing function 
with slope $-B$ until a particular value $x_0$ where the potential drops to zero.  
These two sections are joined at the point of loop pair separation where the reaction coordinate
is equal to $\ell_0$.  For reference, the potential is
\begin{equation}
    U(x) = 
    \begin{cases}
    A x , x < \ell_0 \\
    A \ell_0 - B (x - \ell_0), x > \ell_0.
    \end{cases}
    \label{eq:U-final}
\end{equation}

The average escape time $\langle T \rangle$ for a loop pair of a given size $\ell_0$ 
is then given by the mean first passage time to escape this potential well at $x=0$. Using standard results for 
this problem, we may write
\begin{equation}
    \langle T \rangle = \frac{1}{D} \int_0^{x_0} dy e^{\beta U(y)} \int_0^y dz e^{-\beta U(z)}.
    \label{eq:time-kram}
\end{equation}

The first integral is taken from initial state of the system at $x = 0$ (no loops at all) where the 
potential vanishes $U(0) = 0$, to a so far arbitrary loop separation distance $x_0>0$. 
We return to this point 
shortly. Substituting the potential $U(x)$ from Eq.~\ref{eq:U-final} 
into Eq.~\ref{eq:time-kram} we find that we must perform the integrals
\begin{widetext}
\begin{equation}
    \langle T \rangle = \frac{1}{D} \left( \int_0^{\ell_0} dy  \int_0^y dz e^{\beta A ( y - z)} + \int_{\ell_0}^{x_0} dy \int_0^{\ell_0} dz e^{\beta (A \ell_0 - B (y - \ell_0) - A z)} + \int_{\ell_0}^{x_0} dy \int_{\ell_0}^y dz e^{\beta B ( z - y) } \right).
\end{equation}
\end{widetext}
Doing so, we obtain
\begin{eqnarray}
\langle T \rangle  = \frac{1}{D} \Bigg[   \frac{1}{\beta A }    (\frac{1}{\beta A } ( e^{ \beta A \ell_0} - 1 ) -  \ell_0) 
\\ \nonumber
+
     \frac{1}{\beta A }    \frac{1}{\beta B } (e^{  \beta A \ell_0  } - 1 ) (1 -   e^{  -  \beta B (x_0 - \ell_0) } ) 
\\ \nonumber
+
   \frac{1}{\beta B }  ( x_0 - \ell_0  -   \frac{1}{\beta B } (1  -   e^{- \beta B (x_0 - \ell_0)  } ) \Bigg].
   \label{eq:any-loop-size}
\end{eqnarray}

To determine the upper limit of the above integration, $x_0$, which is the 
final defect separation length, we 
note that, once the potential returns to zero $U(x_0)=0$, the defect pair is now mechanically 
stable and their formation is complete.  Given the slow dynamics of defect motion in 
experiment, one may find that the observed separation 
is, in fact, smaller than this distance if one observes the defects before this mechanical 
equilibrium condition is met. We analyze the above expression in two limiting cases. 
In the first case, we assume that crumpling occurs immediately 
after loop pair production so that $U$ is driven to zero rapidly and the separation distance remains 
small $x_0 \approx \ell_0$.  In the second case, which is expected with small loop production, crumpling will 
not occur until large defect separations so that we may assume $x_0 \gg \ell_0$.

\subsubsection{Large loop approximation}
For the large loop case, we set $x_0 = \ell_0$, and the expression for the escape time is then 
\begin{eqnarray}
\langle T_{\rm large loop} \rangle  = \frac{1}{D} \Bigg[   \frac{1}{\beta A }    (\frac{1}{\beta A } ( e^{ \beta A \ell_0} - 1 ) -  \ell_0)   \Bigg].
\end{eqnarray}

The escape time is minimal (and equal to zero) for the zero loop size. However, this approximation implies 
$\phi \sim 1$, so $\ell_0 \sim \sqrt{\frac{\kappa}{\mu}}$. For low temperature, $\beta \gg 1 / \sqrt{\kappa \mu} $, 
the production of such loops is exponentially suppressed. Thus, we turn to the case of small loops.

\subsubsection{Small loop approximation}
For the small loop case, we expand Eq.~\ref{eq:any-loop-size} and get
\begin{widetext}
\begin{eqnarray}
\langle T \rangle  = \frac{1}{D} \Bigg[ \left(  \frac{1}{\beta A }   \frac{1}{\beta A } e^{ \beta A \ell_0} -   \frac{1}{\beta A }  \frac{1}{\beta A }  -  \ell_0  \frac{1}{\beta A }  \right) 
\\ \nonumber
+
      \frac{1}{\beta A }    \frac{1}{\beta B } e^{  \beta A \ell_0  } - \frac{1}{\beta A }    \frac{1}{\beta B }   -    \frac{1}{\beta A }    \frac{1}{\beta B } e^{  \beta A \ell_0  } e^{  -  \beta B (x_0 - \ell_0) }  + \frac{1}{\beta A }    \frac{1}{\beta B } e^{  -  \beta B (x_0 - \ell_0) }  
\\ \nonumber
+
   \left( \frac{1}{\beta B }   x_0 -   \frac{1}{\beta B }  \ell_0  -    \frac{1}{\beta B }  \frac{1}{\beta B }   +   e^{- \beta B (x_0 - \ell_0)  } \frac{1}{\beta B }  \frac{1}{\beta B }  \right) \Bigg].
   \label{eq:arbitrary-x0}
\end{eqnarray}
\end{widetext}

We now further assume that $B \ll A$, which implies that the compressive force driving the loops apart is 
small compared to the work per unit length necessary to create a loop of a specific size.  In this limit, we 
investigate the case in which the thermal activation of loop pairs should be rather rare.  The opposite limit
of large compressive force implies that loops rapidly proliferate. The ensuing rapid collapse of the bundle 
becomes a different and more complex dynamical problem, which we do not address here. 
Taking into account $B \ll A$ and $\ell_0 \ll x_0$, we get 
\begin{eqnarray}
\langle T \rangle  = \frac{1}{D} \Bigg[         \frac{1}{\beta A }    \frac{1}{\beta B } e^{  \beta A \ell_0  }    
\\ \nonumber
     -    \frac{1}{\beta A }    \frac{1}{\beta B } e^{  \beta (A \ell_0  - B x_0)  }   +   e^{- \beta B x_0  } \frac{1}{\beta B }  \frac{1}{\beta B }    
\\ \nonumber
+
    \frac{1}{\beta B }   x_0   -    \frac{1}{\beta B }  \frac{1}{\beta B }   \Bigg].
\end{eqnarray}
Then in the low temperature limit $e^{- \beta B x_0  } \ll 1$ we get
\begin{eqnarray}
\langle T \rangle  = \frac{1}{D} \Bigg[         \frac{1}{\beta A }    \frac{1}{\beta B } e^{  \beta A \ell_0  }      +    \frac{1}{\beta B }  \left( x_0     -      \frac{1}{\beta B }\right)  \Bigg].
\end{eqnarray} 

The first term does not depend on $x_0$.  It represents the time for the 
pair to be produced, but not for them to 
separate. The second term reflects their overdamped separation dynamics.  The two defects are considered to be 
separated when $x_0 > \frac{1}{\beta B } $ -- the characteristic scale above which their 
recombination is unlikely.  
Thus, we can estimate the bare defect production time  (without taking into account separation) as
\begin{eqnarray}
\langle T_{\rm bare} \rangle  = \frac{1}{D} \Bigg[  \frac{1}{\beta A }    \frac{1}{\beta B } e^{  \beta A \ell_0  }       \Bigg].
\end{eqnarray}
The inverse of $\langle T_{\rm bare} \rangle$ provides the upper bound on the loop pair 
production rate.

To obtain the lower boundary of the production rate, we 
assume the kink angle to remain constant until the 
zero of the potential.  In this case,  the value $x_0$ where the potential again becomes zero is given by 
the linear relation $x_0 = (A + B ) \ell_0 / B $. Substituting this $x_0 $ to Eq.~\ref{eq:arbitrary-x0}, 
we get the equation for the upper boundary of the escape time $\langle T_{\rm upper} \rangle$
that gives the lower boundary for the production rate $r$, which appears in the main text:
\begin{eqnarray}
  \langle T_{\rm upper} \rangle  = \frac{1}{D} \Bigg[         \frac{1}{\beta A }  \frac{1}{\beta A } e^{ \beta A \ell_0} - \frac{1}{\beta A }  \frac{1}{\beta A }   -    \frac{1}{\beta A }  \ell_0  +
\\  \nonumber
+
       \frac{1}{\beta A }    \frac{1}{\beta B }  e^{  \beta A \ell_0  } - 2 \frac{1}{\beta A }    \frac{1}{\beta B }   + e^{  -  \beta  A \ell_0  } \frac{1}{\beta A }    \frac{1}{\beta B }     +
\\ \nonumber
 +
    \frac{1}{\beta B }  A \ell_0 / B -   \frac{1}{\beta B } \frac{1}{\beta B }   + e^{- \beta A \ell_0  } \frac{1}{\beta B } \frac{1}{\beta B } \Bigg]. 
\end{eqnarray}


In the limit of weak forces such that they are just able to 
generate loop pairs, we have $B \ll A$, and consequently we can simplify the mean escape time to
\begin{widetext}
\begin{eqnarray}
 \langle T_{\rm upper} \rangle  = \frac{1}{D} \Bigg[      \frac{1}{\beta^2 A B }     e^{  \beta A \ell_0  } +
     \frac{1}{\beta^2 B^2 }  (\beta A \ell_0    -  1   + e^{- \beta A \ell_0  } ) \Bigg].  
\end{eqnarray}
\end{widetext}
Furthermore, in limit of small angular bends of the loops $\phi$ (consistent 
with small amounts of excess trapped 
length) we approximate: $1 - \cos \phi \approx \phi^2 / 2$. Substituting $A, B$ 
and expressing the total size of the uncross-linked region $\ell_0$ in terms of the loop 
angle using Eq.~\ref{eq:angle-size} we obtain
\begin{eqnarray}
   \langle T_{\rm bare} \rangle  = \frac{1}{D}    \frac{4}{\beta^2 F^2 \phi^4  }  \left(  \frac{ \tau \phi^2 }{2 g_1  }
   e^{  \eta \phi }       +
      \eta \phi    -  1   + e^{- \eta \phi } \right)    
\end{eqnarray}
with $\eta = g_1 g_2 \beta \kappa^{1/2}  \mu^{1/2}  $ and $\tau = F / \mu$.

In the same limit of small loops, we can simplify $T_{\rm bare}$ obtaining
\begin{eqnarray}
   \langle T_{\rm bare} \rangle  =\frac{1}{D} \Bigg[         \frac{1}{\beta g_1 \mu }    \frac{2}{\beta F \phi^2 } e^{ \eta \phi }       \Bigg].   
\end{eqnarray}
We see that the first term in $T_{\rm upper}$ is $T_{\rm bare}$, while other two correspond to the time of 
defect pair separation. Minimizing these times, we find the most likely angle to be produced in 
both approaches. Using $T_{\rm bare}$ and taking a derivative, we find
\begin{eqnarray}
    \phi_{\rm bare} = 2 / \eta.  
\end{eqnarray}

The same calculation for the $T_{\rm upper}$ is a little bit more complicated. To find the 
angle produced with the minimum escape time, we introduce an auxiliary variable  
$\psi = \eta \phi$ in order to write 
\begin{eqnarray}
   \langle T_{\rm upper}  \rangle   = \frac{1}{D} \frac{4 \eta^4}{\beta^2 F^2 \ } f(\psi),
\end{eqnarray}
with 
\begin{eqnarray}
  f(\psi)  = C \psi^{-2}     e^{  \psi }       +
       \psi^{-3}     + (e^{-\psi } - 1 ) \psi^{-4},   
\end{eqnarray}
where $C =  \frac{ \tau  }{2 g_1 \eta^2 }$. For $C \ll 1$ we look for $\psi^\star$ 
such that $f'(\psi^\star) = 0$
in order to extremize the mean escape time. Using the ansatz $\psi \gg 1$, we find that
\begin{equation}
f'(\psi)  \approx C \psi^{-2}     e^{  \psi } - 3 \psi^{-4}.
\end{equation}

From this result, we find the condition for the minimum escape time to be 
\begin{equation}
  C \psi^{2}     e^{  \psi }       = 3,
\end{equation}
which, in the limit $C \ll 1$, has the approximate solution
\begin{equation}
\psi^\star = \log \frac{3}{C} \gg 1.
\end{equation}
This gives the most likely angle to be produced
\begin{equation}
    \phi_{\rm upper} =  \log \frac{3}{C} / \eta. 
\end{equation}

Using the above result for $\psi^\star$, we calculate the minimum escape time. This result gives the lower bound
on the pair production rate of loops that generate the most preferred angle for loop production.  
In other words, we 
expect loops with this angle to be preferentially produced and their production rate will be greater than or 
equal to the following value:
\begin{equation}
    r_{\rm max}  = \frac{1}{\langle T \rangle_{\rm min} }=  
\frac{D}{3} \left[ \frac{ \tau   \log \left(\frac{6 g_1 \eta^2}{\tau} \right)  }{2  \beta \kappa  } \right]^2.
\end{equation}

\section{Braids}

\subsection{Forming a u-bend: Energetics of braid pair production} 
We first consider the energetics of a braid under the compression, using the same approach as 
in Ref.~\cite{Slepukhin:2021}. There we showed that the energy of the most simple 
braid of three filaments can be
mapped onto a two-dimensional projection of the 3D braid onto the plane.  The 2D pseudobraid consists of only 
two filaments.  This simplification results from the observation that in a 
torsion-free braid of filaments in 3D, two
of the three filaments take identical paths (have identical conformations) 
through the defect and maintain their cross linking within that pair, 
while the third takes a different path and loses cross links to the other two.  As a result,
we can lump the energetics of the two identical filaments into one in the analysis of the 2D pseudobraid. We
characterize the conformation of the filament by the angle $\alpha(s)$ between its local tangent and 
a fixed axis, which we take to be the $x$-axis, as a function of arc length $s$

To analyze the pseudobraid, one should consider the case in which the two filaments have differing bending 
moduli.  This makes the analytical solution problematic.  For simplicity, in this analytical part we focus 
only on the case of the equal bending moduli of two filaments in the projection $\kappa_1 = \kappa_2 = \kappa$, 
that presumably results in the symmetric u-bend - two braids with equal angle.  We address the
case of differing bending moduli by direct numerical minimization.   Finally, 
following Ref.~\cite{Slepukhin:2021}, we introduce a symmetry 
ansatz: $\alpha_1(s) = -\alpha_2 (-s) = \alpha(s)$, 
which is justified by later comparing the results with numerical minimization of the energy. 

The boundary conditions are the same as in Ref.~\cite{Slepukhin:2021}
\begin{equation}
\label{eq:angle-bc}
\alpha \left(\pm \frac{L}{2} \right) = \pm \frac{\phi}{2}
\end{equation}
and
\begin{widetext}
\begin{eqnarray}
\label{eq:boundary-condition-braid-sin}
\int_{-L/2}^{L/2} ds  \sin(\alpha(s)) &=& -\int_{-L/2}^{L/2} ds \sin(\alpha(s))  + 2 a \cos \left( \frac{\phi}{2} \right).
\end{eqnarray}
\end{widetext}

The first of these requires the filaments to once again be parallel at the ends of the braid defect and ensure 
that the tangent rotates through the full kink angle as the arc length variables moves through the braid
defect. The second enforces their normal separation of $a$ as fixed by the cross-linking 
molecules at each end of the braid defect.

In addition to this previous analysis of braids in bundles with free ends, the 
energy from Ref.~\cite{Slepukhin:2021} now needs a contribution from the compression $F R \cos \phi$:
\begin{widetext}
\begin{eqnarray}
\nonumber
E  &=& 2 \Bigg[
\lambda_s \left[\int_{-L/2}^{L/2} \! \! \! \! \! ds  \sin(\alpha(s)) -  \! \! \! \! \!   -  a \cos\left( \frac{\phi}{2} \right) \right] 
+ \\
 &+& \int_{-L/2}^{L/2} \! \! \! \! \! ds \kappa \frac{\alpha'^2}{2} + \tilde{\lambda} \left[ \delta \left( s + \frac{L}{2} \right) \alpha + \frac{\phi}{ 2 L} \right] + \lambda^\dagger \left[ \delta \left(s - \frac{L}{2} \right) \alpha \nonumber 
 - \frac{\phi}{ 2 L} \right]    \Bigg] \\  &+&  \mu L + F R  \cos  \phi,
\label{eq:energy}
\end{eqnarray}
\end{widetext}
where $\lambda$, $\tilde{\lambda}$ ,$\lambda^\dagger $, and $\lambda_s $  are Lagrange multipliers 
fixing boundary conditions given in Eqs.~\ref{eq:angle-bc} and \ref{eq:boundary-condition-braid-sin}.  

We vary the energy Eq.~\ref{eq:energy} with respect to $\alpha(s)$, obtaining the equation of 
elastic equilibrium, which is the 
same as the one obtained in Ref.~\cite{Slepukhin:2021} for the braid with free boundaries: 
\begin{eqnarray}
\label{Eq-for-alpha1}
\kappa \alpha &=& \lambda_s \cos \alpha_1 +  \tilde{\lambda}_1\delta \left( s + \frac{L_1}{2} \right)+ \lambda_1^{\dagger} \delta \left( s - \frac{L_1}{2} \right).
\end{eqnarray}
Eq.~\ref{Eq-for-alpha1} requires us to impose the equalities:  $\tilde{\lambda}= \kappa \alpha'\left( -\frac{L}{2} \right)$ and $\lambda^\dagger = -\kappa \alpha' \left( \frac{L}{2} \right) $.
Repeating arguments from~\cite{Slepukhin:2021} we show that  $\lambda_s$ must
be negative everywhere.   Minimizing the energy Eq.~\ref{eq:energy}  with respect to the length of the defect 
$L$ we obtain
\begin{eqnarray}
\label{Result-of-varying-L1}
 \int_{-L/2}^{L/2}  ds \left(- \frac{\kappa}{2} \alpha'^2  +  \mu + \lambda_c \cos \alpha + \lambda_s \sin \alpha \right) &=& 0. 
\end{eqnarray}
Using Eq. \ref{Eq-for-alpha1},  we find a conserved quantity (the analog of first integrals of a 
dynamical system) $I$ such that $dI  / ds  = 0$.  This first integral is given by 
\begin{eqnarray}
\label{Conserv1}
 I &=&  \frac{\kappa}{2} \alpha'^2 -  \lambda_s \sin  \alpha. 
\end{eqnarray}
Comparing Eq.~\ref{Conserv1} and Eq.~\ref{Result-of-varying-L1} we find the value of this 
first integral, which is
the binding energy per unit length of the cross linkers:
\begin{equation}
\label{Integral-of-motion}
  \frac{\kappa}{2} \alpha'^2 -  \lambda_s \sin  \alpha = \mu.
\end{equation}
From the result the rate of change of the tangent angle at the ends of the braid defect are fixed.  We find
\begin{equation}
\alpha'^2\left(\pm \frac{L}{2} \right) = 2 \left[ \frac{\mu}{\kappa} \pm \lambda_s \sin  \left(\frac{\phi}{2} \right) \right]. 
\label{angle-sqr}
\end{equation}
Subtracting in the Eq.~\ref{angle-sqr} the upper sign equation from the lower sign one, we obtain 
a helpful expression for later use:
\begin{eqnarray}
\label{Ends-correspondence1}
  \frac{4 \lambda_s \sin  \left(\frac{\phi}{2} \right)}{\alpha'\left(\frac{L}{2} \right) -  \alpha'\left( - \frac{L}{2} \right)} =  \kappa \left[ \alpha'\left(\frac{L}{2} \right) +  \alpha'\left(-\frac{L}{2} \right) \right]. 
\end{eqnarray}

Finally, the variation of the total energy Eq.~\ref{eq:energy} with respect to $\phi$ gives
\begin{equation}
\frac{\partial E}{\partial \phi}   = 4 \lambda_s   a \sin \left( \frac{\phi}{2} \right)
 + 2 \tilde{\lambda}   -2  \lambda^\dagger - F R \sin \left(\phi \right).
\label{eq:variation}
\end{equation}
Using boundary condition Eq.~\ref{Eq-for-alpha1}, we simplify Eq.~\ref{eq:variation} to
\begin{widetext}
\begin{equation}
\frac{\partial E}{\partial \phi}   =  4 \lambda_s   a \sin \left( \frac{\phi}{2} \right)
 + \kappa \alpha'\left(-\frac{L}{2} \right) + \kappa \alpha'\left(-\frac{L}{2} \right) + \kappa \alpha'\left(\frac{L}{2} \right)+ \kappa \alpha'\left(\frac{L}{2} \right)  - F R \sin \left(\phi \right).
 \label{variation-2}
\end{equation}
\end{widetext}

Substituting  Eqs.~\ref{Ends-correspondence1} into Eq.~\ref{variation-2} we find
\begin{equation}
\frac{\partial E}{\partial \phi}   =  4 \lambda_s    \sin \left( \frac{\phi}{2} \right)\left[  a  +
   \frac{4 }{\alpha'\left(\frac{L}{2} \right) -  \alpha'\left(-\frac{L}{2} \right)}   +  \frac{F R}{2 \lambda_s} \cos(\phi/2) \right].
   \label{var-phi}
\end{equation}
Eq.~\ref{var-phi} shows that $\phi = 0$ is an extremum.The transition from 
minimum to maximum occurs when the expression in square brackets is zero: 
\begin{equation}
  a  +
   \frac{4 }{\alpha'\left(\frac{L}{2} \right) -  \alpha'\left(-\frac{L}{2} \right)} -  \frac{F R }{2 \lambda_s} \cos(\phi/2) = 0.
 \label{extremum-sym}
\end{equation}
While this transition from minimum to maximum indicates the second order phase transition, we also may 
have a situation in which there is another potential minimum at $\phi \neq 0 $. If, at some value of 
system parameters, this minimum becomes deeper than the minimum at $\phi = 0$ (even when it is still minimum, 
not maximum), we obtain a first order phase transition instead of the second order one. As we see using 
numerical energy minimization, the symmetric case corresponds to the second order phase transition, 
while the asymmetric produces a first-order kinking transition. Here we continue to study the symmetric case
to analytically explore the second order phase transition.

Let us consider the critical point at which the braided bundle transitions from an unkinked state with 
$\phi=0$ to one with finite kinking angles.  Assuming the transition is second order so that the kinking angle 
grows continuously from zero at the transition, we may study this point by first setting 
$\phi = 0$. Eq.~\ref{angle-sqr} then transforms to 
\begin{equation}
\alpha'\left(\pm \frac{L}{2} \right) = \pm \sqrt{ 2 \frac{\mu}{\kappa} }.
\end{equation}
Using it, we simplify Eq.~\ref{extremum-sym} to
\begin{equation}
a  -   \frac{2}{ \sqrt{2 \mu/\kappa}}  =  \frac{F R}{2 \lambda_s}.  
\label{force-lambda}
\end{equation}
\begin{figure*}
    \centering
    \includegraphics[width=\linewidth]{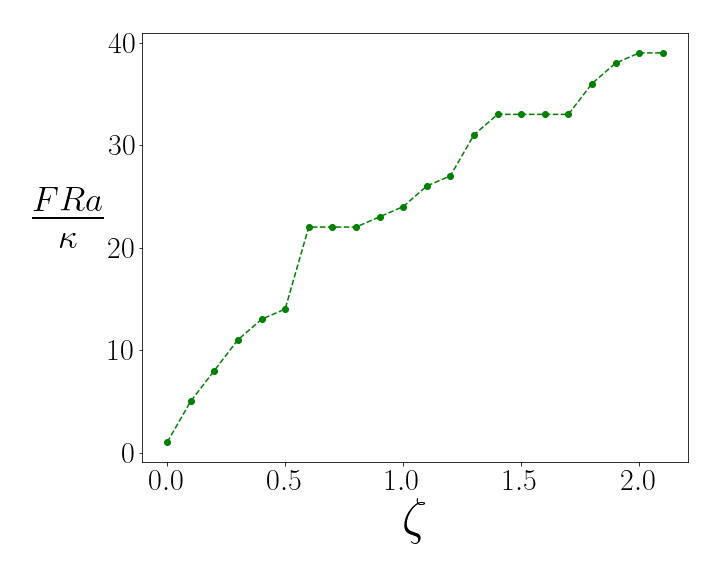}
    \caption{The compressive load required to produce Euler buckling in a three-filament
    bundle as a function of $\zeta$. Green dots - results of numerical minimization of the energy. Dashed line - guide for the eye.}
    \label{fig:threshold-si}
\end{figure*}


As we move through the braid, $\alpha$ increases from 0 to its maximal value $\alpha_{\rm max}$ and then 
decreases back to 0. $\alpha_{\rm max}$ is determined by $\alpha'_{\rm max} = 0$, so 
Eq.~\ref{Integral-of-motion} gives
\begin{equation}
 \mu =   -\lambda_s \sin  \alpha_{\rm max}, 
\end{equation}
and
\begin{equation}
\alpha_{\rm max} = \arcsin ( - \mu / \lambda_s ).  
\end{equation}

Due to symmetry of the problem the maximum angle occurs at the origin of the arc length variable:
$\alpha_{\rm max} = \alpha(0)$.  The integral condition Eq.~\ref{eq:boundary-condition-braid-sin} 
for $\phi = 0$ gives
\begin{equation}
\int_{-L/2}^{L/2} ds \sin \alpha = a.
\end{equation}
Due to symmetry, the integral can be transformed to 
\begin{equation}
\int_{-L/2}^0 ds \sin \alpha = a/2,
\label{eq:half-integral}
\end{equation}
where $\alpha$ grows from $0$ to $\alpha_{\rm max}$. 
Denoting $z = \sin \alpha(s)$ so that $\alpha (s) = \arcsin z$, and 
\begin{equation}
\alpha' (s) ds  = \frac{dz}{\sqrt{1 -z^2} }. 
\end{equation} 
Then Eq.~\ref{Integral-of-motion} gives
\begin{equation}
\alpha'  = \pm \sqrt{2 (\mu  + \lambda_s z)  /  \kappa}.   
\end{equation}
Then on the interval under consideration, $\alpha$ grows so 
\begin{equation}
ds  =  \frac{dz}{\sqrt{1 -z^2} \sqrt{2 (\mu  + \lambda_s z)  /  \kappa}  }. 
\end{equation} 
Substituting this identity into the integral in Eq.~\ref{eq:half-integral} we may write
\begin{equation}
\int_0^{ -\mu / \lambda_s }  \frac{z dz}{\sqrt{1 -z^2} \sqrt{2 (\mu  + \lambda_s z)  /  \kappa}  }  = a/2.
\end{equation}
Finally, we use Eq.~\ref{force-lambda} to determine $\lambda_s$ and after some some algebra, 
we obtain an equation relating compressive load and $\zeta$ at critical point
\begin{equation}
\int_0^1  \frac{\left( \left( \sqrt{2 \zeta} - \zeta \right)  
\frac{2 \kappa } {F R a} \right)^2 t dt}{\sqrt{1 - \left( \left( \sqrt{2 \zeta} - \zeta \right)  
\frac{2 \kappa } {F R a} \right)^2 t^2} \sqrt{1  - t)}  }  
= \sqrt{\zeta/2}.
\label{final}
\end{equation}
This implicit relation between between $\zeta$ and the critical compressive load $F^\star = F^\star(\zeta)$
can be solved numerically.  The results are shown in Fig.3 of the main text. 

\section{Euler buckling of the bundle}

We determine the critical force for buckling of a three-filament bundle 
by numerically minimizing its energy:
\begin{equation}
E = - F \Delta L + \mu \ell + \sum_{i=1}^N  \int ds  \frac{\kappa_i}{2} \left(\partial_s \hat{t}_i \right)^2.
\label{eq:general-energy-in-SI}
\end{equation}
In this minimization, we do not insert either braid or loop defects, 
but the bundle when bent does 
loose cross linkers.  This uncross-linked region may be viewed as some sort of localized 
defect, but it is not one of the defects whose annealing requires a system-sized 
rearrangement of cross links.

We observe that, for a strong enough compression, the state where the central part of the
bundle bends has less energy than the straight bundle. The size of the bent part $\ell$ is
determined by the minimization of the energy Eq.~\ref{eq:general-energy-in-SI}. We find the
threshold compression value at which bundles transform from straight to bent, as a function
of $\zeta$ (see Fig.~\ref{fig:threshold-si} SI). We observe that this value is
significantly bigger than the braid-induced kinking threshold (see Fig.~3 main document),
showing that defect production occurs before buckling upon increasing force.

\section{Group theory of braids and loops}

A defect on a three-filament bundle consists of the combination of braiding of the filament and the trapping of
excess length of these filaments. Braiding is characterized by the braid group, which was described in relation 
to the current problem in Ref.~\cite{Slepukhin:2021}. Here we repeat the main points of these discussion.

We number the filaments (1,2,3) such that (1) is on the left, (3) is on the right, (2) is on the middle. 
It is easy to see that, after the defect their order may change, for example the filament that 
used to be (1) becomes filament (3). 

The standard braid group with 3 strands has two generators: $\sigma_1$ (corresponding to passing strand 2 
over the strand 3), and $\sigma_2$ (corresponding to passing strand 3 over strand 1). 
We also introduce an operator $\sigma_3$, corresponding to passing strand 1 over strand 2, which can be 
written in terms of the other two generators and their inverses: 
$\sigma_3 = \sigma_1 \sigma_2^{-1} \sigma_1^{-1}$.  

\begin{figure*}
\centering
\includegraphics[width=0.8\linewidth]{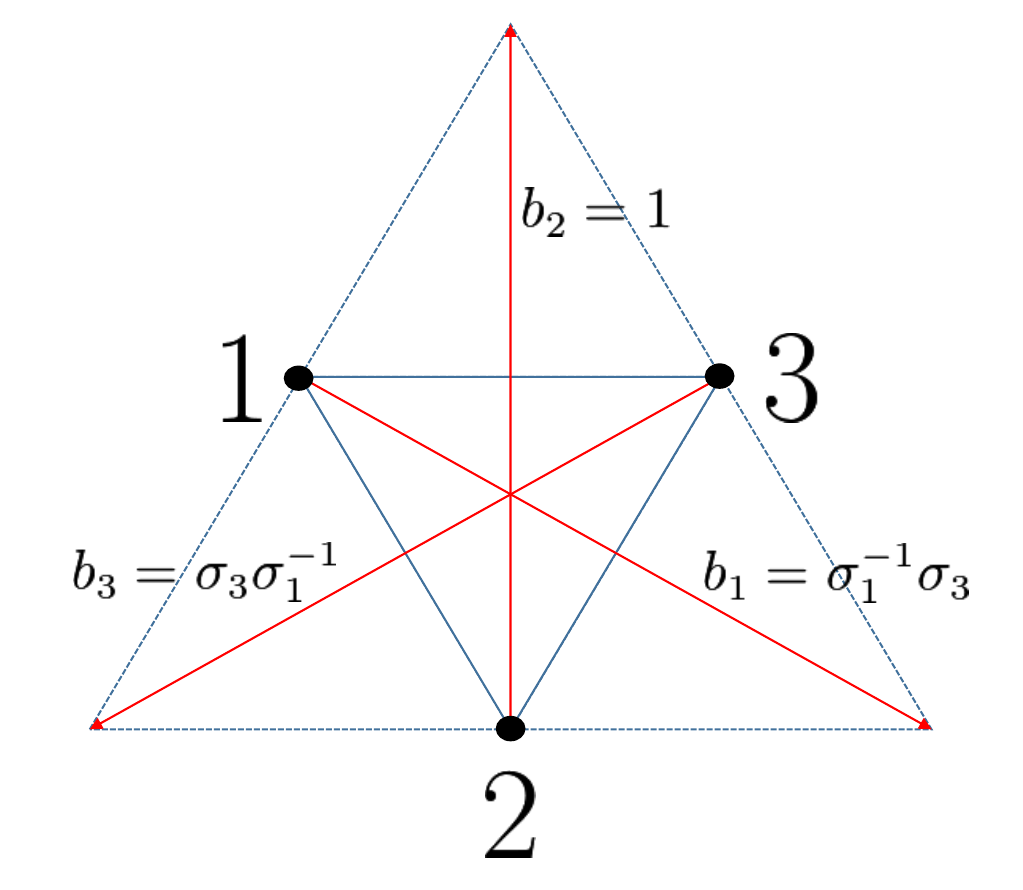}
\caption{Illustration of the braiding operators $b_{1,2,3}$ acting on a bundle with filaments 1,2,3 in a state with 
filament 2 below the plane, where braiding operator $b_i$ corresponds to the braiding of the $i^{\rm th}$ 
filament. For example, $b_1$ corresponds to first passing 
filament 1 {\em over} filament 2,
which is done by operator $\sigma_3$. After that, the former filament 1 is now filament 2, and vice versa. 
The second and final step in the braid operation is passing the new filament 2 (i.e., old filament 1) 
{\em under} filament 3, which is performed by operator $\sigma_1^{-1}$. We obtain the formulae for 
other operators in the same way. The figure is reused from Ref.~\cite{Slepukhin:2021}} 
\label{fig:braid-group}
\end{figure*}

The group of interest, however,  is not identical to the braid group describing the 
filament bundle. Indeed, the braid group does not take into 
account the 3D nature of the problem.  If we take a projection of the bundle onto the plane containing 
filaments 1 and 3, filament 2 can either be 
above or below this plane (see Fig.~\ref{fig:braid-group}). Those are two different states of the 
bundle that are 
not distinguished by the standard braid group. To distinguish them, we introduce additional $Z_2$ group that 
denotes whether the filament 2 is above or below the plane. 
Then the total group controlling the state of the bundle due to braiding (and twists) is the direct product  
of $Z_2$ and the braid group: $\mbox{Br}_3 \times Z_2 $.


We also need to take into account the excess length trapped in the filaments. There are three filaments with 
possibly three separate excess 
lengths $\Delta L_{1,2,3}$.  Only their differential excess lengths matter for characterizing the defects. These
differences are defined by two numbers: $\Delta L_1 - \Delta L_2$ and $\Delta L_2 - \Delta L_3$. The lengths 
are elements of the the group of reals under addition $\mathbb{R}$.
Thus, a defect is totally characterized by an element of direct product $\mbox{Br}_3 \times Z_2 \times \mathbb{R}^2$. 
In the simplest braids discussed here, we consider the case where two filaments making up the braid have no length 
difference, but the third carries excess length. Moreover, that excess length is set by energy minimization to obtain the
lowest energy braid.  Clearly more complex braids cost higher energy, and can be thought of as being composed of combinations of 
braids and loops. In principle, loops (with zero braid group charge) too may carry more than one 
excess length associated with their different filaments.  We consider here the case of loops in which only one filament carries 
excess length.  In that case, the charge associated with the loop is a scalar.  In more complex loops having multiple filaments looping
from the bundle, that scalar charge should be thought of as a vector of such charges corresponding to the excess lengths of the various 
constituent filaments. Such structures can be thought of a composite objects made up of multiple simple loops.

\section{Phase diagram for braid/loop production}
\begin{figure*}
    \centering
    \includegraphics[width = 15cm]{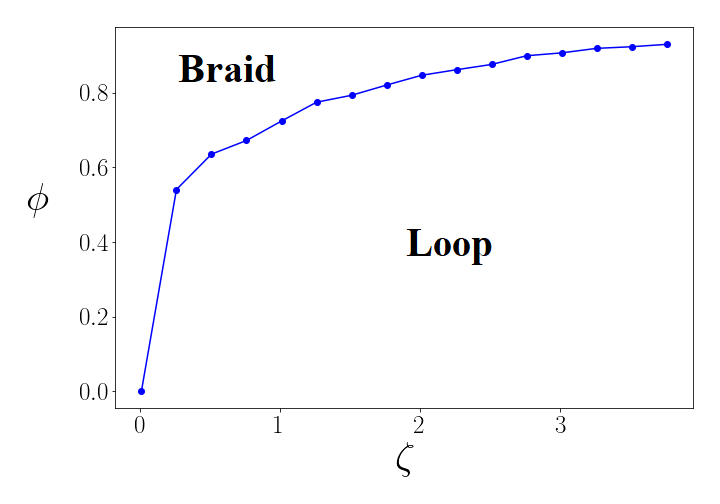}
    \caption{At a fixed torque of $\frac{F R a}{\kappa} = 2.1$, 
    loops producing kink angles less than $\phi(\zeta)$ are energetically favored over 
    braid formation.  Similar curves (not shown) can be produced for varying torques.}
    \label{fig:phase-boundary-fixed-torque}
\end{figure*}

To explore whether loops or braids are more likely to be produced, we formulate the 
question in the following way. 
Given the value of compressive load and material parameters defining $\zeta$, we find the 
energy required to produce a braid pair (with excess length consistent with the minimum of this energy), 
or a pair of loops for various values of excess length. We choose the simple criterion 
that the lower energy configuration
is more likely to be produced. Since braids can form a u-bend, compression decreases their energy 
at the moment of formation; loops, however, form only z-bends, so the compression almost does not effect 
them at the moment of formation. Instead, the 
shortening the bundle relies on the loop separation, forming z-bend.  
Therefore, increasing compression enhances the likelihood of braid formation over loop formation.
For loops, the energy of the loop (as well as the initially produced angle) 
increases with the amount of trapped length. 
If we specify the value of the torque and material properties $\zeta$, all loops produced with an 
angle less than a certain threshold are more likely to be produced than any braid. 
We show this threshold angle as a function of the dimensionless parameter $\zeta$ for a particular 
value of the torque $\frac{F R a}{\kappa} = 2.1$ in the Fig.~\ref{fig:phase-boundary-fixed-torque}.   

\begin{figure*}
    \centering
    \includegraphics[width = 15cm]{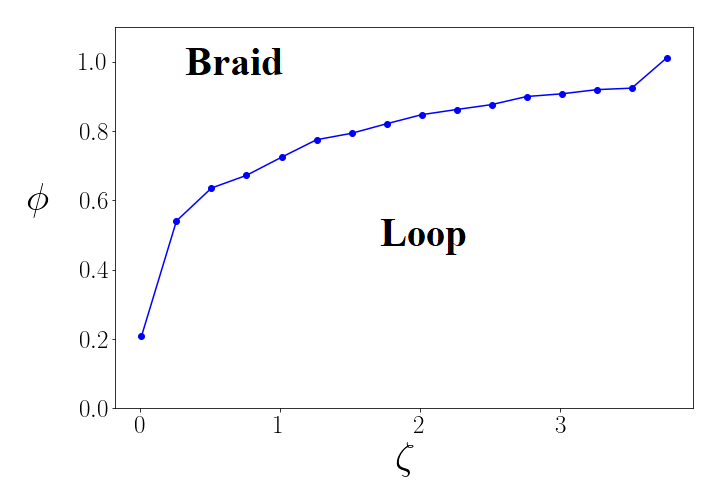}
    \caption{Loops with initial kink angles less than $\phi(\zeta)$
    are energetically favored over the formation braids with kink angle $\pi/2$. The applied torque is adjusted with 
    changing $\zeta$ in order to fix the braid angle.}
    \label{fig:phase-boundary-fixed-angle}
\end{figure*}

For small values of $\zeta$ and large enough torques, bundles crumple, folding at the 
defects into large-angle kinks. The braid formation is precluded by Euler buckling in a 
finite region of the bundle that is now free of cross linkers. 
To avoid this effect at very small $\zeta$, we consider only those values of torque 
such that each braid produces the angle $\pi/2$. Of course, then there are different torque 
values for each $\zeta$. For such fixed-angle braids, we obtain the threshold angle below which 
loop production will dominate over those particular braids, which is 
shown in Fig.~\ref{fig:phase-boundary-fixed-angle}.

\bibliography{bib_kei, bib_bundles}

\end{document}